# On the Structure of the Yang-Mills Vacuum

G. Schierholz

*Deutsches Elektronen-Synchrotron DESY*
*Notkestraße 85, 22603 Hamburg, Germany*

and

*Gruppe Theorie der Elementarteilchen, Höchstleistungsrechenzentrum HLRZ*
*c/o Forschungszentrum Jülich, 52425 Jülich, Germany*

e-mail: gsch@x4u2.desy.de

ABSTRACT

In this talk I will discuss the current picture of color confinement. In particular, I will show how it can be tested microscopically. It is stressed that the color magnetic monopoles in this picture are dyons. Furthermore, the role of instantons is illuminated.

## 1. Introduction

Color confinement is certainly the most striking phenomenon of the low-energy behavior of QCD. Twenty years ago 't Hooft [1] and Mandelstam [2] have conjectured that color confinement could be understood in terms of a dual superconductor in which color magnetic monopoles condense and color electric charges are confined through a dual Meissner effect. Eight years ago this idea was formulated [3] and successfully tested for the first time [4] on the lattice. Since then we have made steady progress in deciphering the mechanism of color confinement. For a review of the most recent developments I refer the reader to the talks of Di Giacomo, Haymaker and Suzuki at this conference.

It seems that the idea of monopole condensation is in essence correct. It must be said, however, that we still know very little about the dynamics that drives it. To improve our knowledge I find it necessary to look at the problem from as many 'angles' as possible. In this talk I shall take the opportunity to present some new results in this direction.

---

## 2. Instantons or Monopoles?

One of the characteristic features of the Yang-Mills vacuum is the existence of instantons. They give rise to an integer valued topological charge

$$Q = -\frac{1}{32\pi^2} \int d^4x \, \epsilon_{\mu\nu\rho\sigma} \text{Tr} F_{\mu\nu} F_{\rho\sigma} \in \mathbb{Z}. \tag{1}$$

A classical instanton corresponds to a field configuration of minimal action with $Q = 1$ which is self-dual, i.e.

$$F_{\mu\nu} = \tilde{F}_{\mu\nu} \equiv \frac{1}{2}\epsilon_{\mu\nu\rho\sigma} F_{\rho\sigma}. \tag{2}$$

Instantons represent tunneling events between vacua of different winding number $n$, $|n\rangle$. They cause the vacuum state $|n\rangle$, including the perturbative vacuum $|0\rangle$, to become quantum mechanically unstable. The proper vacuum states, which are stable under any gauge invariant operation, are

$$|\theta\rangle = \sum_n \exp(i\theta n)|n\rangle. \tag{3}$$

These so-called $\theta$ vacua are realized by adding a CP violating term to the action,

$$S \to S_\theta = S - i\theta Q, \tag{4}$$

where $S$ is the standard action. By convention I will take $\theta$ to lie in the interval $[0, 2\pi)$. A priori $\theta$ is a free parameter. Since no CP violation has been observed in the strong interactions, $\theta$ must however be very close to zero. This is known as the strong CP problem.

By fixing the gauge so that only the gauge degrees of freedom of the Cartan subgroup $U(1)^{N-1} \subset SU(N)$ are kept dynamical, the gluonic degrees of freedom of the theory may be mapped onto 'photons', color electric charges and color magnetic charges, i.e. monopoles [5]. The idea is that the long-distance physics is essentially described by the abelian degrees of freedom, which is called abelian dominance.

Let me, for the sake of later use, be a little more explicit. The abelian vector potentials $a^i_\mu, i = 1, \ldots, N$, refered to as 'photons', are taken to be the diagonal components of the (original) $SU(N)$ gauge potentials after gauge fixing. They form abelian field strengths

$$f^i_{\mu\nu} = \partial_\mu a^i_\nu - \partial_\nu a^i_\mu, \tag{5}$$

which lead to the definition of magnetic currents

$$K^i_\mu = \frac{1}{8\pi} \epsilon_{\mu\nu\rho\sigma} \partial_\nu f^i_{\rho\sigma}. \tag{6}$$

We recall that integration of the current density over a three-dimensional region $\Omega$ yields

$$m^i(\Omega) = \int_\Omega d^3\sigma_\mu K^i_\mu = \frac{1}{8\pi} \int_{\partial\Omega} d^2\sigma_{\mu\nu} \epsilon_{\mu\nu\rho\sigma} f^i_{\rho\sigma}. \tag{7}$$

Taking $\Omega$ at constant time ($\mu = 4$) we obtain

$$m^i(\Omega) = \frac{1}{4\pi} \int_{\partial\Omega} d^2\sigma_j b^i_j, \quad b^i_j = \frac{1}{2}\epsilon_{jkl} f^i_{kl}, \tag{8}$$

which is the magnetic flux through $\partial\Omega$ and hence counts the magnetic charge inside $\Omega$. The magnetic charges obey the Dirac quantization condition $m^i = 0, \pm\frac{1}{2}, \pm 1, \ldots$. (In our notation the quark would have color electric charge 1.) An elementary monopole has charge $m^i = \frac{1}{2}$.

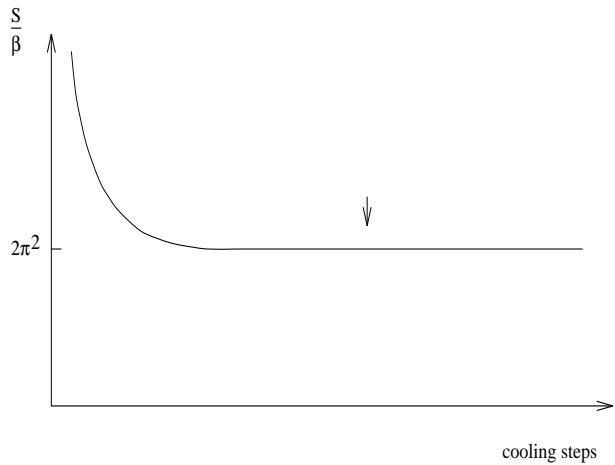

Fig. 1. Cooling history for gauge group $SU(2)$ and a single instanton event. The action for this event is $S/\beta = 2\pi^2$.

It is widely believed [6] that instantons cannot cause confinement. This is probably true for a dilute instanton gas. But for an instanton liquid, for example, the situation could be different [7]. Naively one would expect that instantons consist of closed dyon loops. Bornyakov and I have looked into this question [8]. We generated a set of lattice instantons by cooling $SU(2)$ gauge field configurations [9], as schematically shown in Fig. 1. We then brought these configurations into the maximally abelian gauge and performed the abelian projection. The result was a monopole loop located at the core of the instanton as I have shown in Fig. 2. For larger instantons (and larger correlation lengths) we would expect to find also larger monopole loops. Thus instantons are a source for color magnetic monopoles. If the instantons overlap, as they do in the liquid, it might be possible that the monopoles form a plasma, which then could lead to color confinement. Later on I will show that these monopoles are in fact dyons.

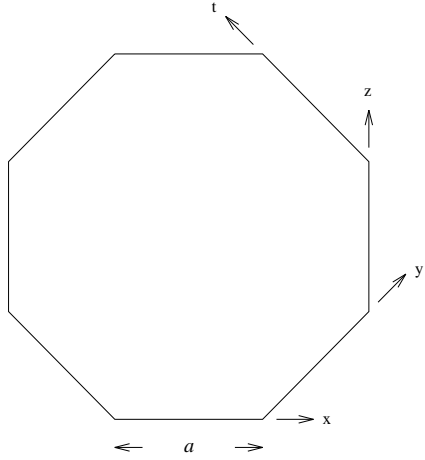

Fig. 2. Perspective view of a monopole loop induced by instantons. At $\beta = 2.3$ and on a $8^4$ lattice the loops extended typically over one or two lattice spacings $a$.

This means that there is not only one vacuum which we have to understand but there are infinitely many. The $\theta$ vacua may be vastly different [10], or not. That will depend on the dynamics of instantons and monopoles: at non-zero values of $\theta$ the monopoles will be exposed to a background instanton field. Our investigation of instanton events [8] indicates that instantons and monopoles are closely knotted together. I would therefore expect that instantons play an essential role in the formation of the Yang-Mills ground states. What role they play will be revealed in particular by exploring the whole spectrum of Yang-Mills vacua.

## 3. $\theta$ Vacua and Dyons

In the $\theta$ vacuum the color magnetic monopoles acquire a color electric charge of the magnitude [11]

$$e^i(\theta) = \frac{\theta}{2\pi} \, 2. \tag{9}$$

In our notation [3] gluons have color electric charge two which is the origin of the factor 2 on the right hand side of eq. (9). This means that in the $\theta$ vacuum the monopoles become dyons. In Fig. 3 I have plotted the color electric and magnetic charges of the abelian degrees of freedom, and as they evolve as $\theta$ is varied from 0 to $2\pi$. We see that at $\theta = 2\pi$ the monopole-gluon bound state has taken over the position of the monopole. Because the theory is periodic in $\theta$ with period $2\pi$, at least one phase transition is required. 't Hooft has argued [5] that one phase transition occurs at $\theta \approx \pi$, where the monopole may bind together with the monopole-gluon bound state of opposite charge and form a condensate. In this phase, the so-called oblique confinement phase, quarks may escape disguised as bosons by picking up a monopole from the vacuum. Ezawa and Iwazaki [12] even argued that there are infinitely many phase transitions.

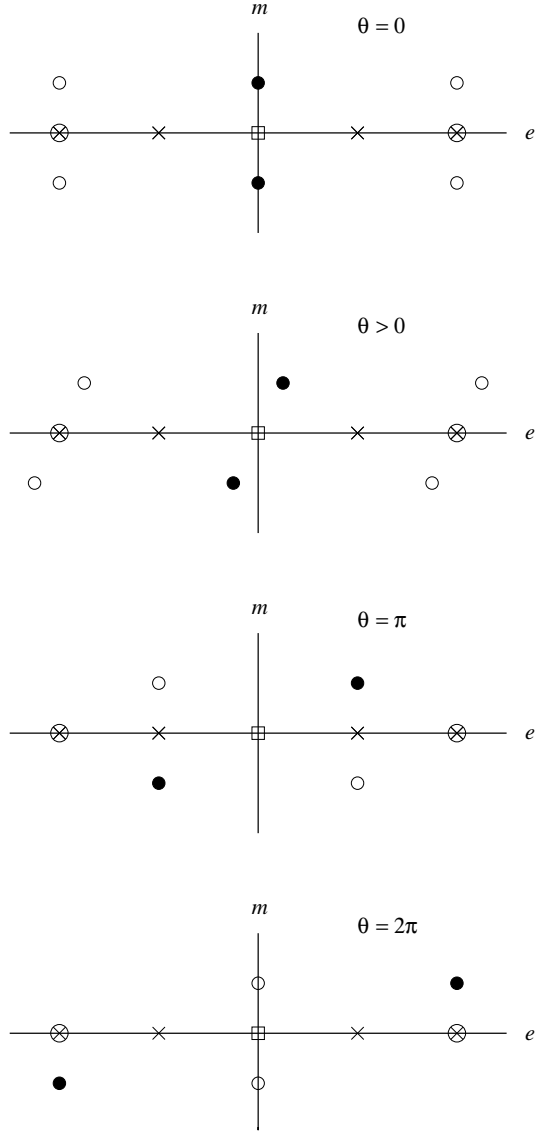

Fig. 3. The electric ($e$) and magnetic ($m$) charge lattice for various values of $\theta$. The solid circles represent monopoles, crosses represent quarks, crossed circles gluons, open circles monopole-gluon bound states, and the box represents the 'photon'.

The $\theta$ dependence of the vacuum is given by the partition function

$$Z(\theta) = \sum_Q \exp(i\theta Q)p(Q) \equiv \exp(-VF(\theta)), \qquad (10)$$

where $p(Q)$ is the probability of finding a field configuration with topological charge $Q$, and $F(\theta)$ is the free energy per space-time volume $V$. Let me now discuss the consequences of eq. (9). I define the average (minkowskian) color electric charge of the dyon in the background of a field configuration with topological charge $Q$ by $e^i(Q)$. It then follows that

$$\frac{\theta}{\pi} Z(\theta) = i \sum_Q \exp(i\theta Q)p(Q)e^i(Q). \qquad (11)$$

Solving this equation for $e^i(Q)$, one obtains

$$e^i(Q) = \frac{1}{\pi} \sum_{\bar{Q} \neq Q} \frac{(-1)^{\bar{Q}-Q}}{\bar{Q} - Q} \frac{p(\bar{Q})}{p(Q)}. \qquad (12)$$

It is easy to see that $e^i(0) = 0$ and $e^i(-Q) = -e^i(Q)$. Thus, another interpretation of eq.(9) is that color magnetic monopoles turn into dyons in the background of an instanton field configuration. If the monopole travels through the background of an instanton–anti-instanton pair configuration we would furthermore conclude that it changes its charge at some stage from (say) positive to negative value by radiating off gluons. And so on. Finally, on a large lattice with a large number of instantons and anti-instantons we would expect to find dyons with frequently fluctuating charges.

To show that this picture is basically correct, at least for the maximally abelian gauge, let me return to the classical instanton discussed in the last section. The color electric charge of the monopole is given by

$$e^i(\Omega) = \frac{1}{2\pi} \int_\Omega d^3 \sigma_\mu \partial_\nu f^i_{\mu\nu}. \qquad (13)$$

Taking $\Omega$ at constant time ($\mu = 4$), this gives

$$e^i(\Omega) = \frac{1}{2\pi} \int_{\partial\Omega} \mathrm{d}^2\sigma_j e^i_j, \ \ e^i_j = f^i_{4j}. \tag{14}$$

We have computed [8] $e^i$ for the monopoles in Fig. 2 that originated from the instanton. Not only did we find that the charge density had its maximum exactly at the position of the monopole, but also that all eight (anti-)monopoles had approximately charge $+1$ for an instanton and charge $-1$ for an anti-instanton, provided the three-dimensional region $\Omega$ was chosen large enough. Thus, we would conclude that the dyon charge is quantized, contrary to the classical dyon solution [13].

This is not completely surprising. A classical (anti-)instanton configuration obeys $F_{\mu\nu} = \pm\frac{1}{2}\epsilon_{\mu\nu\rho\sigma}F_{\rho\sigma}$, with $+$ sign for the instanton and $-$ sign for the anti-instanton. If this (anti-)self-duality survives the abelian projection, which eventually it should if the long-distance physics is indeed described by the abelian degrees of freedom, then we would have $f^i_{\mu\nu} = \pm\frac{1}{2}\epsilon_{\mu\nu\rho\sigma}f^i_{\rho\sigma}$. If we now insert this into eq. (14), we obtain from eq. (8)

$$e^i(\Omega) = \pm 2m^i(\Omega) = \pm 1, \tag{15}$$

which is what we found numerically. This result may also be taken as a nice confirmation of the hypothesis of abelian dominance.

What consequences does this have for the dynamics of the Yang-Mills vacuum? In the confining vacuum we would expect dyons to be confined. This would mean that color confinement is only consistent with zero vacuum angle $\theta$. Moreover, it suggests that large instantons and anti-instantons are expelled from the confining vacuum. Smaller instantons and anti-instantons about the size of a correlation length or smaller, which sum up to $Q = 0$, would probably be allowed. They are certainly needed for a solution of the $U_A(1)$ problem.

What makes a lattice simulation of $\theta$ vacua very difficult is the fact that the action is complex for non-vanishing values of $\theta$. In order to facilitate things, I have first considered the CP$^3$ model [14]. The CP$^{N-1}$ models in two space-time dimensions are in many respects similar to QCD. The common properties include (i) the existence of instantons and a vacuum angle $\theta$, (ii) asymptotic freedom, (iii) a dynamically generated mass gap, (iv) dimensional transmutation and (v) confinement. In two dimensions confinement is not difficult to achieve. But in the CP$^{N-1}$ models confinement arises without the presence of any fundamental gauge field.

In Fig. 4 I show the free energy $F(\theta)$ on a $64^2$ lattice at $\beta = 2.7$. The correlation length at this value of $\beta$ is $\approx 9$. We observe a kink at $\theta = \theta_c \approx 0.5$. This is a signal for a first order phase transition in $\theta$. What happens dynamically can be read off

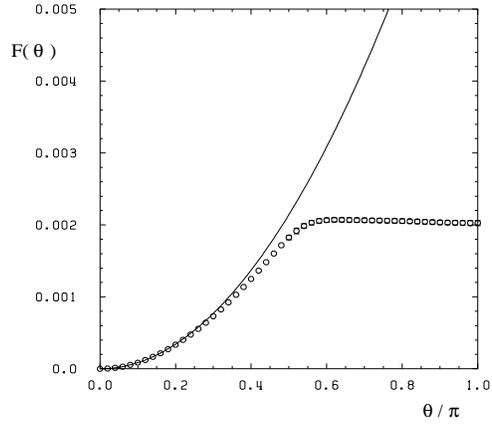

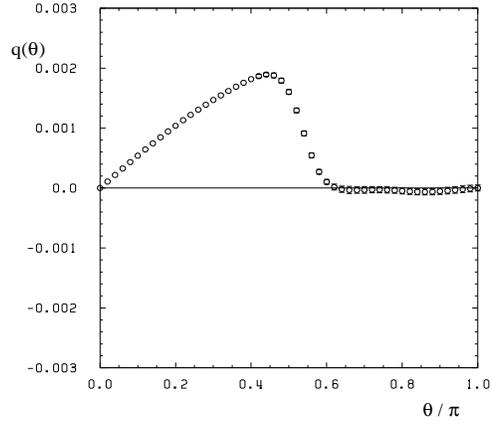

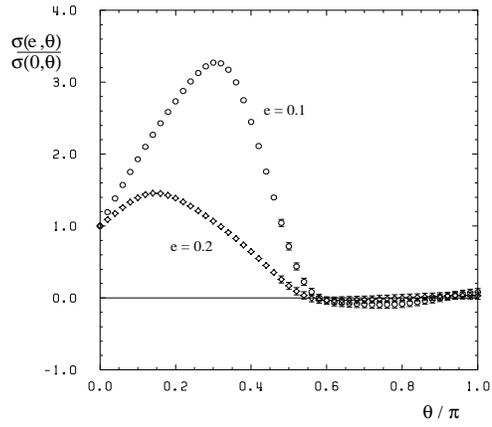

Fig. 4. The free energy $F(\theta)$, the topological charge density $q(\theta)$ and the string tension $\sigma(e,\theta)$ as a function of $\theta$ for the $CP^3$ model on the $64^2$ lattice at $\beta = 2.7$. The solid curve in the figure at the top is the prediction of the large-N expansion. In the string tension figure at the bottom $e$ denotes the external charge in units of the fundamental charge.

from the topological charge density

$$q(\theta) = \frac{\mathrm{d}F(\theta)}{\mathrm{d}\theta}, \qquad (16)$$

which in this case equals the average electric field induced by the instantons. I show this quantity also in Fig. 4. We see that $q(\theta)$ increases almost linearly up to $\theta = \theta_c$ where it jumps to zero and then stays zero. This indicates that the phase transition is driven by the collapse of the background electric field. Finally I have computed the string tension as a function of $\theta$. This is also plotted in Fig. 4. We find that the string tension for properly chosen external charges is non-zero for $\theta < \theta_c$ while it vanishes for $\theta > \theta_c$. Thus the theory deconfines for $\theta > \theta_c$.

We can expect to find that the theory deconfines for all $\theta > 0$ only in the continuum limit, i.e. for $\beta \to \infty$. To see this happen I have done calculations at various values of $\beta$. The result of the calculations is summarized by the phase diagram in Fig. 5. It turns out that the critical value of $\theta$, $\theta_c$, decreases towards zero as $\beta$ goes to infinity. The horizontal lines marked by arrows are lines of constant physics. For $\theta > 0$ these lines end on the line of first order phase transitions. Only the line with $\theta = 0$ will allow the cut-off to be taken to infinity. This would also solve the strong CP problem if there is no new physics that crosses the way.

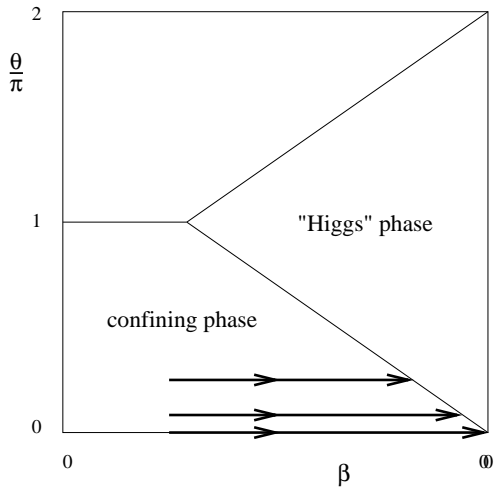

Fig. 5. The phase diagram. The line which starts at $\theta = \pi$ in the strong coupling region and then turns towards $\theta = 0$ (and $\theta = 2\pi$, respectively) in the continuum limit is a line of first order pase transitions. The horizontal lines marked by arrows are lines of constant physics.

Can this serve as a model for QCD? I think yes, because it is primarily the background electric field which causes the conflict with confinement.

In four dimensions the problem is computationally voluminous and progress is very slow. Brandstaeter and I have investigated the SU(2) Yang-Mills theory. We use Phillips and Stone's algorithm [15] for the topological charge. We find basically the same behavior as in the case of the $CP^3$ model: at small values of $\theta$ the free energy increases proportional to $\theta^2$, it then shows a kink indicating a first order phase transition, and finally it flattens off to a constant. First results of our calculation were reported in Ref. 16. The next step is to map out the phase diagram. For that we have to repeat the calculation for several other values of $\beta$.

## 4. Conclusions

The results I have presented here are a first step towards a microscopic understanding of the Yang-Mills vacuum. Many of the intuitive ideas we have developed in the last years seem to prove to be correct finally. So, most likely, we are on the right track.

The main conclusion of this talk is probably that there is no sharp division between instantons and monopoles. We need the features of both to explain the properties of the Yang-Mills vacuum.

## 5. Acknowledgements

I like to thank Vitaly Bornyakov and Frank Brandstaeter for discussions and collaboration on the subject of this talk.